\documentstyle[multicol,prb,aps,psfig]{revtex}
\begin{document}
\draft
\title{Light Scattering from Nonequilibrium Concentration
Fluctuations in a Polymer Solution}
\author{W.B.~Li\footnote{Present address: @Road, Inc., 45635 Northport
Loop East, Freemont, CA 94538.}, K.J.~Zhang\footnote{Present
address: Electronic Data Systems, 800 K Street NW, Washington, DC
20001.}, J.V.~Sengers, and R.W.~Gammon\\ {\it Institute for
Physical Science and Technology, University of Maryland,\\ College
Park, MD 20742, USA}\\ J.M.~Ortiz de Z\'{a}rate\\ {\it Facultad de
Ciencias F\'{\i}sicas, Universidad Complutense, 28040 Madrid,
Spain}}
\date{\today}
\maketitle

\begin{abstract}
We have performed light-scattering measurements in dilute and
semidilute polymer solutions of polystyrene in toluene when
subjected to stationary temperature gradients. Five solutions with
concentrations below and one solution with a concentration above
the overlap concentration were investigated. The experiments
confirm the presence of long-range nonequilibrium concentration
fluctuations which are proportional to $(\nabla T)^2/k^4$, where
$\nabla T$ is the applied temperature gradient and $k$ is the wave
number of the fluctuations. In addition, we demonstrate that the
strength of the nonequilibrium concentration fluctuations, observed
in the dilute and semidilute solution regime, agrees with
theoretical values calculated from fluctuating hydrodynamics.
Further theoretical and experimental work will be needed to
understand nonequilibrium fluctuations in polymer solutions at
higher concentrations.
\end{abstract}


\begin{multicols}{2}
\narrowtext

\section{Introduction}

Density and concentration fluctuations in fluids and fluid mixtures
can be investigated experimentally by light-scattering techniques.
The nature of these fluctuations when the system is in equilibrium
is a subject well understood\cite{equili,equili2}. Here we shall
consider fluctuations in nonequilibrium steady states (NESS), when
an external and constant temperature gradient is applied, while the
system remains in a hydrodynamically quiescent state. That is, we
shall deal with fluctuations that are intrinsically present in
thermal nonequilibrium states in the absence of any convective
instabilities. Such fluctuations have received considerable
attention during the past decade\cite{chemphys}. It was originally
believed that, because of the existence of local equilibrium in
NESS, the time correlation function of the scattered-light
intensity would be the same as in equilibrium, but in terms of
spatially varying thermodynamic and transport properties
corresponding to the local value of temperature. However, it has
been demonstrated that qualitative differences do appear. The first
complete expression for the spectrum of the nonequilibrium
fluctuations in a one-component fluid subjected to a stationary
gradient was obtained by Kirkpatrick et al.\cite{kirk1} by using
mode-coupling theory. They showed that the central Rayleigh line of
the spectrum would be substantially modified as a result of the
presence of a temperature gradient. Because of a coupling between
the temperature fluctuations and the transverse-velocity
fluctuations through the temperature gradient, spatially long-range
nonequilibrium temperature and viscous fluctuations appear,
modifying the Rayleigh spectrum of the scattered-light intensity.
Their results were subsequently confirmed on the basis of
fluctuating hydrodynamics\cite{ronis,law1}. The effect is largest
for the transverse-velocity fluctuations in the direction of the
temperature gradient which corresponds to the situation that the
scattering wave vector, ${\bf k}$, is perpendicular to the
temperature gradient $\nabla T$, which configuration will be
assumed throughout the present paper. In that case, the strengths
of the nonequilibrium temperature and viscous fluctuations are
predicted to be proportional to $(\nabla T)^2/k^4$. The dependence
on $k^{-4}$ implies that, in real space, the nonequilibrium
correlation functions become long ranged\cite{chemphys,real}. The
spatially long-range nature of the correlation functions in NESS is
nowadays understood as a general phenomenon arising from the
violation of the principle of detailed balance\cite{chemphys,long}.
Experimentally, the long-range nature of the nonequilibrium
fluctuations can be probed by light-scattering measurements at
small wave numbers $k$, i.e. at small scattering angles $\theta$.
Such experiments have been performed in one-component liquids and
excellent agreement between theory and experiments has been
obtained\cite{reva,physica}.

In binary systems, the situation is a little more complicated. In
liquid mixtures or in polymer solutions a temperature gradient will
induce a concentration gradient through the Soret effect. This
induced concentration gradient is parallel to the temperature
gradient and has the same or opposite direction depending on the
sign of the Soret coefficient, $S_T$. In this case, nonequilibrium
fluctuations appear, not only because of a coupling between the
temperature fluctuations and the transverse-velocity fluctuations
through the temperature gradient, but also because of a coupling
between the concentration fluctuations and the transverse-velocity
fluctuations through the induced concentration gradient. The
nonequilibrium Rayleigh-scattering spectrum has been calculated for
binary liquid mixtures, both on the basis of mode-coupling
theory\cite{nieu} and on the basis of fluctuating
hydrodynamics\cite{theory}, with identical results. In liquid
mixtures in thermal nonequilibrium states not only nonequilibrium
temperature and nonequilibrium viscous fluctuations, but also
nonequilibrium concentration fluctuations exist. The strengths of
all three types of nonequilibrium fluctuations are again predicted
to be proportional to $(\nabla T)^2/k^4$. The theory was extended
by Segr\`{e} et al.\cite{gravity,physica2} to include the effects
of gravity and by Vailati and Giglio\cite{giglio2} to include
time-dependent nonequilibrium states. The nature of the
nonequilibrium fluctuations in the vicinity of a convective
instability has also been investigated\cite{physica2,cross,cohen},
in which case the nonequilibrium modes become propagative. The
influence of boundary conditions, which may become important when
${\bf k}$ is parallel to $\nabla T$, was considered by
Pagonabarraga et al.\cite{rubi}.

Experiments have been performed to study nonequilibrium
fluctuations in liquid mixtures\cite{physica,approx,cell,giglio}.
The three types of nonequilibrium fluctuations have been observed
in liquid mixtures of toluene and
n-hexane\cite{physica,approx,cell} and the strength of all three
types of nonequilibrium fluctuations were indeed proportional to
$(\nabla T)^2/k^4$, as expected theoretically. Initially, it seemed
that also the prefactors of the amplitudes of these nonequilibrium
fluctuations were in agreement with the theoretical
predictions\cite{approx}. However, a definitive assessment was
hampered by a lack of reliable experimental information on the
Soret coefficient\cite{approx,cell}. To our surprise, subsequent
accurate measurements of the Soret coefficient of liquid mixtures
of toluene and n-hexane obtained both by K\"{o}hler and
M\"{u}ller\cite{kohler} and by Zhang et al.\cite{kai1} yielded
values for the Soret coefficient that were about 25\% lower than
the values needed to explain the quantitative magnitude of the
amplitudes of the observed nonequilibrium fluctuations\cite{kai1}.
Subsequent measurements of the nonequilibrium concentration
fluctuations in a mixture of aniline and cyclohexane, obtained by
Vailati and Giglio\cite{giglio} did not have sufficient accuracy to
resolve this issue.

As an alternative approach, we decided to investigate
nonequilibrium concentration fluctuations in a polymer solution. In
a polymer solution the same nonequilibrium enhancement effects are
expected to exist, but the mass-diffusion coefficient, $D$, in this
case is several orders of magnitude lower than in ordinary liquid
mixtures. In addition, the Soret coefficient is two orders of
magnitude larger that the Soret coefficient of ordinary liquid
mixtures\cite{kaithesis}. These two facts, as discussed below,
simplify the theory because, as also happens in an equilibrium
polymer solution, the concentration fluctuations become dominant
and they are readily observed by light scattering. Both the data
acquisition and the data analysis become much easier in this case.
Thus a polymer solution would seem to be an ideal system to further
investigate nonequilibrium concentration fluctuations.

For this purpose we have selected solutions of polystyrene in
toluene, for which reliable information on the thermophysical
properties is available. We have performed small-angle
Rayleigh-scattering experiments in polystyrene-toluene solutions
subjected to various externally applied temperature gradients. A
summary of our results has been presented in a Physical Review
Letter\cite{letter}. In the present paper we provide a full account
of the experiment and of the analysis of the experimental data.

\section{Theory}

A theory specifically developed for the fluctuations in polymer
solutions in thermal nonequilibrium states is not yet available in
the literature. However, for polymer solutions in the dilute and
semidilute solution regime, as long as entanglement effects can be
neglected, it should be possible to use the same fluctuating
hydrodynamics equations as those for ordinary liquid mixtures. As
mentioned in the introduction, the complete expression for the
Rayleigh spectrum of a binary liquid in the presence of a
stationary temperature gradient was evaluated by Law and
Nieuwoudt\cite{theory}. The dynamic structure factor contains three
diffusive modes. The decay rate $\lambda_{\nu}=\nu k^2$ of one of
these modes is determined by the kinematic viscosity $\nu$; this
mode disappears in thermal equilibrium, i.e. when $\nabla T
\rightarrow 0$. The other two modes are also present in the
Rayleigh spectrum of a liquid mixture in equilibrium and have the
decay rates $\lambda_{\pm}$ given by:
\begin{equation}
\lambda_{\pm} = {k^2 \over 2} (D_T + {\cal D}) \mp
{k^2 \over 2} \left[ (D_T + {\cal D})^2 - 4 D_T D \right]^{1/2},
\label{eq:lamda}
\end{equation}
where $D_T$ is the thermal diffusivity, $D$ the binary
mass-diffusion coefficient, in the case of a polymer solution to be
referred to as collective diffusion coefficient\cite{kaithesis},
and where
\begin{equation}
{\cal D}=D(1+\epsilon)
\end{equation}
with
\begin{equation}
\epsilon = T~{[S_T ~ w(1-w)]^2 \over c_{P,c}}
~ \left( {\partial \mu \over \partial w} \right)_{p,T}.
\label{eq:epsilon}
\end{equation}
In Eq. (\ref{eq:epsilon}) $w$ is the concentration expressed as
mass fraction of the polymer, $c_{P,c}$ the isobaric specific heat
capacity of the solution at constant concentration, and $\mu$ is
the difference between the chemical potentials per unit mass of the
solvent and the solute. As usual, the Soret coefficient, $S_T$,
specifies the ratio between temperature and concentration gradients
and it is defined through the steady-state phenomenological
equation:
\begin{equation}
\nabla w = - S_T~w~(1-w)~\nabla T.
\label{eq:soret}
\end{equation}

The two modes with decay rates $\lambda_{\pm}$ incorporate a
coupling between temperature and concentration fluctuations. This
coupling becomes important in compressible fluid mixtures near a
critical locus\cite{anisimov}. However, when $\epsilon \ll 1$, the
dynamic structure factor of a binary system in NESS consists of
just three exponentials with decay rates given by:
\begin{equation}
\lambda_{\nu}=\nu k^2, ~ \lambda_{+}=D_T k^2, ~ \lambda_{-}=D k^2.
\label{eq:decay}
\end{equation}
Physically, it means that the temperature gradient $\nabla T$
induces nonequilibrium viscous fluctuations appearing as a new term
in the Rayleigh spectrum and it leads to nonequilibrium
enhancements of the temperature and concentration fluctuations.

The condition $\epsilon \ll 1$ is always fulfilled in the
low-concentration limit\cite{equili2}. But for some liquid mixtures
$\epsilon$ is also small at all concentrations. For instance, in an
equimolar mixture of toluene and n-hexane Segr\`{e} et al. found
$\epsilon \approx 0.028$. For the dilute polymer solutions
considered in the present paper, $\epsilon \approx 0.010$. Hence,
the approximations implying the presence of three diffusive modes
with the simple decay rates given by Eq. (\ref{eq:decay}) are even
more justified for a polymer solution than for the liquid mixtures
studied in previous papers\cite{physica,approx}.

Furthermore, in ordinary polymer solutions $D/D_T \approx
5\times10^{-4}$; such a small value simplifies the analysis of the
concentration fluctuations, because the decay in the time
correlation function coming from the concentration mode (decay rate
$\lambda_{-}$) will be well separated from the decays of the modes
with decays rates $\lambda_{\nu}$ and $\lambda_+$. Actually, for
the small angles employed in our experiments, the decay time of the
$\lambda_{-}$ mode is typically 0.5-1.5 s, whereas the  decay times
of the other two modes are around $10^{-4}-10^{-5}$ s. In addition,
for a dilute polymer solution, the Rayleigh factor ratio, that
determines the ratio of the scattering intensities of the
concentration fluctuations and the temperature
fluctuations\cite{approx}, is much larger than unity so that the
contributions of temperature fluctuations are negligibly small in
practice. Moreover, when the experimental correlograms are analyzed
beginning at $10^{-2}$ s, viscous and temperature fluctuations have
already decayed. This kind of approximation is usually assumed in
the theory of light scattering from polymer solutions in
equilibrium states\cite{equili}.

In conclusion, for heterodyne light-scattering experiments in
sufficiently dilute polymer solutions, the expression for the
time-dependent correlation function of the scattered light becomes:
\begin{equation}
C(k,t) = C_0[1+A_c({\bf k},\nabla T)] e^{-D k^{2} t},
\label{eq:corr}
\end{equation}
where $C_0$ is the signal to local-oscillator background ratio
representing the amplitude of the correlation function in thermal
equilibrium. In Eq. (\ref{eq:corr}) the term $A_c({\bf k},\nabla
T)$ represents the nonequilibrium enhancement of the concentration
fluctuations. This term is anisotropic and depends on the
scattering angle. When ${\bf k} \perp \nabla T$, $A_c$ reaches a
maximum given by:
\begin{equation}
A_c(k,\nabla T) = A^*_{c}(w)~{(\nabla T)^2 \over k^4},
\label{eq:enhan1}
\end{equation}
where the strength of the enhancement, $A^*_{c}(w)$, is given
by\cite{physica,theory,physica2,approx}:
\begin{equation}
A^*_{c}(w) = {\left[ w (1-w) \right]^{2} {S_T}^2 \over \nu D} ~
\left( {\partial \mu \over \partial w} \right)_{p,T} ~
\left[ 1 + 2 {D \over D_T} (1 + \zeta) \right],
\label{eq:enhan11}
\end{equation}
where $\zeta$ is a dimensionless correction term related to the
ratio of $(\partial n / \partial T)_{P,w}$ and $(\partial n /
\partial w )_{P,T}$. As mentioned earlier for the polymer solutions
to be considered, $D/D_T$ is of the order of $5 \times 10^{-4}$, so
that the correction term inside the square brackets can be
neglected in practice. Hence, the expression for $A^*_{c}(w)$
reduces to:
\begin{equation}
A^*_{c}(w) = {\left[ w (1-w) \right]^{2} {S_T}^2 \over \nu D} ~
\left( {\partial \mu \over \partial w} \right)_{p,T}.
\label{eq:enhan2}
\end{equation}

The dependence of the nonequilibrium enhancement $A_c$ of the
concentration fluctuations on $k^{-4}$ indicates that the
correlations in real space are long ranged. Actually, a $k^{-4}$
dependence in Fourier space corresponds to a linear increase of the
correlation function in real space \cite{real}. The rapid increase
of the strength of the nonequilibrium fluctuations with decreasing
values of the wavenumber will saturate for sufficiently small $k$
due to the presence of gravity. The gravity effect was predicted by
Segr\`{e} et al.\cite{gravity} and has been confirmed by some
beautiful experiments of Vailati and Giglio\cite{giglio}.

Our set of working equations, Eqs. (\ref{eq:corr}),
(\ref{eq:enhan1}) and (\ref{eq:enhan2}), can be also deduced from
the theory of Vailati and Giglio for nonequilibrium fluctuations in
time-dependent diffusion processes\cite{giglio2}. In that paper
only concentration and velocity fluctuations are considered around
a nonequilibrium time-dependent state. By applying an inverse
Fourier transform to Eq. (25) of Vailati and Gilio\cite{giglio2},
we obtain for the autocorrelation function of the concentration
fluctuations for ${\bf k} \perp \nabla w$:
\begin{equation}
\langle \delta w \delta w^{*} \rangle = S(k)
\exp \left[-Dk^2t \left(1-{R(k) \over R_c} \right) \right],
\label{eq:giglio1}
\end{equation}
where $R(k)/R_c$ is the Rayleigh-number ratio, which accounts for
the effects of gravity and is defined in Eq. (22) of Vailati and
Giglio\cite{giglio2}. $S(k)$ is the static structure factor,
defined in Eq. (26) of the same paper. It should be noticed that,
for consistency, in our Eq. (\ref{eq:giglio1}) we have used the
symbol $w$ instead of the $c$ used by Vailati and Giglio. Both
symbols have the same meaning, as it is stated after Eq. (2) of
Vailati and Giglio that: ''$c$ is the weight fraction of the denser
component". Thus Eq. (26) of Vailati and Giglio for the static
structure factor $S(k)$ can be written as:
\end{multicols}
\widetext
\begin{equation}
S(k) = {k_B T \over 16 \pi^4 \rho} ~ \left( {\partial w \over
\partial \mu} \right)_{p,T} ~ {1 \over 1 - R(k)/R_c} ~ \left[ 1 +
{(\nabla w)^2 \over \nu D k^4} ~ \left( {\partial \mu \over
\partial w} \right)_{p,T} ~ \right].
\label{eq:giglio2}
\end{equation}
\begin{multicols}{2}
\narrowtext
The derivation of Vailati and Giglio\cite{giglio2} was originally
developed for time-dependent isothermal diffusion processes in the
presence of gravity, where only concentration gradients are
present. The validity of this theory for stationary nonequilibrium
states is obvious, and we can consider $\nabla w$ to be constant,
neglecting the weak dependence on space and time considered by
Vailati and Giglio\cite{giglio2}. Neglecting gravity effects is
equivalent to assuming that the Rayleigh-number ratio is almost
zero, $R(k)/R_c \approx 0$, as can be easily shown from Eq. (22) in
the paper of Vailati and Giglio\cite{giglio2}. Introducing Eq.
(\ref{eq:soret}) into Eq. (\ref{eq:giglio2}) and neglecting
$R(k)/R_c$, one readily verifies that Eqs. (\ref{eq:giglio1}) and
(\ref{eq:giglio2}) are equivalent to Eqs. (\ref{eq:enhan1}) and
(\ref{eq:enhan2}).

We note that Eqs. (\ref{eq:corr}), (\ref{eq:enhan1}) and
(\ref{eq:enhan2}) can also be obtained as a limiting case of the
equations derived by Schmitz for nonequilibrium concentration
fluctuations in a colloidal suspension \cite{schmitz}. Schmitz
considered a colloidal suspension, in the presence of a constant
gradient, $\nabla \phi$, in the volume fraction, $\phi$, of the
colloidal particles, maintained against diffusion by continuous
pumping of solvent between two semipermeable and parallel walls.
The relevant expression is Eq. (7.7) in the article of
Schmitz\cite{schmitz}, which gives the dynamic structure factor,
$S(k,\omega)$, of the nonequilibrium concentration fluctuations as
a function of the wavenumber $k$ and the frequency $\omega$. The
expression obtained by Schmitz includes non-local and memory
effects due to the large particle sizes in comparison with the
scattering wavelength. The non-local and memory effects cause the
transport coefficients to depend on the frequency and the wave
number. In our case these effects can be neglected, because the
polymer molecules are much smaller that the colloidal particles
considered by Schmitz and the condition $kR_{\rm g} \ll 1$, where
$R_{\rm g}$ is the radius of gyration, is fulfilled. With this
simplification the original expression obtained by Schmitz reduces
to:
\begin{equation}
S(k,\omega) = S(k) ~ [1+A_c] ~ {2 D k^2 \over \omega^2 + (D
k^2)^2},
\end{equation}
where $A_c$ is given by:
\begin{equation}
A_c = {w^2 \over \nu D} ~ \left( {\partial \mu \over \partial w}
\right)_{p,T} ~
\left( {\nabla \phi \over \phi} \right)^{2}~ {1 \over k^4} .
\label{eq:schmitz2}
\end{equation}
In deriving Eq. (\ref{eq:schmitz2}) we have used the relationship
between pumping rate and volume fraction gradient, as given by Eqs.
(7.6) and (7.11) in the paper of Schmitz\cite{schmitz}. We have
also converted the various concentration units used by Schmitz: $n$
which is the number of particles per unit volume ($n=\rho /
w~m_{\rm p}$) and $c$ which is the number of particles per unit
mass ($c = w/m_{\rm p}$), $\rho$ being the density of the
suspension and $m_{\rm p}$ the mass of one particle. Furthermore,
we are using the difference between solvent and solute chemical
potentials per unit mass, whereas in the paper of Schmitz $\mu$ is
the difference in chemical potentials per particle.

Although the theory of Schmitz for the nonequilibrium concentration
fluctuations was derived for the case that the concentration
gradient is produced by a solvent flow, the results will be
applicable to NESS under a constant temperature gradient. Since the
solvent flow does not appear explicitly in Eq. (\ref{eq:schmitz2}),
we may also apply the equation to a system subjected to a
stationary volume-fraction gradient caused by other driving forces,
such as a temperature gradient. For our polymer solutions $(\nabla
\phi / \phi)^2  \approx (\nabla w / w)^2$, the difference being
less than 0.5\%. With this approximation and by using Eq.
(\ref{eq:soret}), it is straightforward to show the equivalence of
the theory for nonequilibrium concentration fluctuations in a
colloidal suspension and in a polymer solution, independent of
whether the concentration gradient is established by solvent
pumping or induced by a temperature gradient.

\section{Experimental Method}

The polystyrene used to prepare the polymer solutions was purchased
from the Tosoh Corporation (Japan); it has a mass-averaged
molecular weight $M_W = 96,400$ and a polydispersivity of 1.01, as
specified by the manufacturer\cite{kaithesis}. The solvent toluene
was Fisher certified reagent purchased from Baker Chemical Co. with
a stated purity of better than 99.8\%. By using the correlation
proposed by Noda et al.\cite{noda1,noda2}, $\langle R_{\rm g}^2
\rangle = 1.38 \times 10^{-2} M_W^{1.19}$, the radius of gyration,
$R_{\rm g}$, and the overlap concentration, $w^*$, for this polymer
in toluene solutions may be estimated as $R_{\rm g} \approx 11.0$
nm and $w^* \approx 3.1\%$, respectively. Therefore, for the
viscoelastic and entanglement effects to be negligible, the
concentrations employed in this work should not be much higher than
$3.1\%$. Five polystyrene/toluene solutions below $w^*$ and one
slightly above $w^*$ were prepared gravimetrically as described by
Zhang et al.\cite{kaithesis}. To assure the homogeneity of the
mixture, the solutions were agitated (magnetic stirrer or shaking
by hand)for at least one hour before further use.

\begin{figure}
\[
\psfig{file=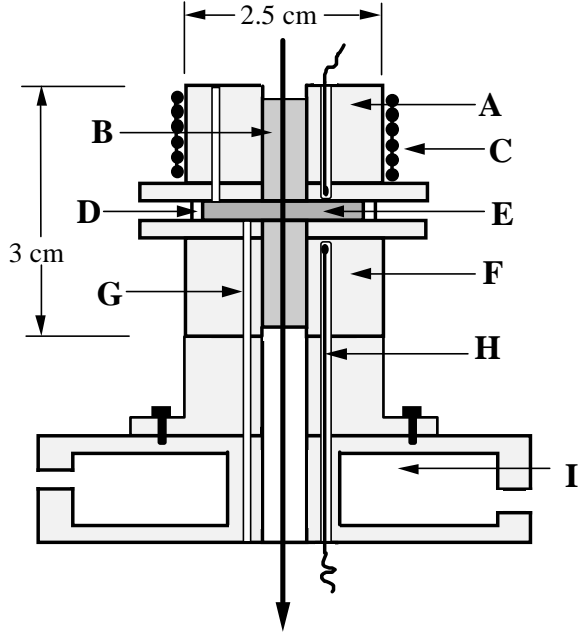,width=8.5cm}
\]
\caption{Schematic representation of the optical cell for small-angle
Rayleigh scattering. A: top plate. B: sapphire windows. C:
resistive heater. D: fused quartz ring. E: sample liquid layer. F:
bottom plate. G: filling tubes. H: thermistors. I: water
circulation chamber.}
\label{f1}
\end{figure}

The light-scattering experiments were performed with an apparatus
specifically designed for small-angle Rayleigh scattering in an
horizontal fluid layer subjected to a vertical temperature
gradient\cite{physica}. A diagram of the optical cell is shown in
Fig.\ \ref{f1}. The following paragraphs will describe the
apparatus in detail, and references to the symbols in Fig.\
\ref{f1} will be used. The polymer solution in the actual
light-scattering cell (E) is confined by a circular quartz ring (D)
that is sandwiched between top and bottom copper plates (A and F,
respectively). The cell is filled through two stainless steel
capillary tubes (G) which had been soldered into holes in the top
and bottom plates. The inner and the outer diameters of the quartz
ring (D) are 1.52 cm and 2.05 cm, respectively. Two identical
optical windows (B) are used for letting a laser beam pass through
the liquid solution. Both windows are cylindrical and made from
sapphire because of the relatively high thermal conductivity of
this material. The windows are epoxied into the centers of the top
and the bottom copper plates by a procedure similar to the one
described by Law et al.\cite{window}. Once the optical windows are
installed, the top and bottom copper plates are sealed against the
quartz ring by indium O-rings. To avoid heat conduction from the
upper to the lower plate other than through the liquid, the two
copper plates are held together by teflon screws (not shown in the
figure). The tension of these teflon screws was adjusted to set the
flat surfaces of the optical windows parallel to within 10 seconds
of arch. This was accomplished by passing a laser beam through the
cell and monitoring the interference pattern of the reflected beams
from the inner surfaces of the windows while the screws were
fastened. Once the cell was mounted, the distance $d$ between the
windows was accurately determined by measuring the angular
variation of these interference fringes. The result was $d = (0.118
\pm 0.005)$ cm. This value was confirmed by also measuring the
separation with a cathetometer.

Since the nonequilibrium enhancement of the fluctuations given by
Eq. (\ref{eq:enhan1}) depends on $k^{-4}$, small scattering angles
are required to obtain measurable values of $A_c$; in practice
scattering angles $\theta$ between 0.4$^{\circ}$ and 0.9$^{\circ}$
were used. A major difficulty with such small-angle experiments is
the  presence of strong static scattering from the optical
surfaces. To reduce the background scattering, we needed very clean
windows. The optical surfaces of the windows and the other optical
components of the experimental arrangement were cleaned by applying
first a Windex solution and then acetone with cotton swabs.
Moreover, we employ thick cell windows to remove the air-window
surfaces from the field of view of the detector. In addition, the
outer surface of the windows and the other optical surfaces were
broad-band (488-623 nm) anti-reflection coated (CVI Laser
Corporation), to reduce the back reflections and forward scattering
intensity. As a result of these efforts, we obtained high enough
signal to background ratios to allow accurate measurements of the
scattered-light intensities.

To observe the intrinsic nonequilibrium fluctuations, it is
essential to avoid any convection in the liquid layer. This is
accomplished by heating the horizontal fluid layer from above.
Furthermore, bending and defocusing of the light beam due to the
refractive-index gradient induced by the temperature gradient could
cause serious limitations in the resolution of the angles. This
difficulty is avoided by employing a vertical incident light beam,
parallel to the temperature gradient. To eliminate stray
deflections caused by air currents near the cell, the whole
assembly is covered by a plexiglass box with a small hole at the
top to let the laser beam pass through.

The temperature at the top plate was maintained by a computer
controlled resistive heater winding (C). The temperature at the
bottom plate is controlled by circulating constant-temperature
fluid from a Forma Model 2096 bath through channels in the base of
the cell (I), which is in good thermal contact with the bottom
plate. With these devices, the temperatures of both the hot and
cold plates can be held constant to within $\pm$20 mK. These
temperatures remain fixed over the collection time of any
experimental run. Temperatures of both plates are monitored by
thermistors inserted in holes (H) next to the sample windows and
deep enough to be located very close to the liquid layer. Due to
the relatively high thermal conductivity of the sapphire windows
and the symmetry of the arrangement, the lateral temperature
gradients are small. Numerical modeling of the thermal conduction
process yields negligible differences between the measured
temperature gradient and the actual temperature gradient between
the windows.

A coherent beam ($\lambda = 632.8$ nm) from a 6 mW stabilized cw
He-Ne laser is focused with a 20 cm focal length lens onto the
polymer solution in the scattering cell. The scattering angle,
$\theta$ is selected with a small pinhole (500 $\mu$m) located 197
mm after the cell. The collecting pinhole is placed in a plane
orthogonal to the transmitted beam. The distance between the point
where the beam hits the collection plane and the pinhole was
carefully measured with a vernier micrometer scale.

In the original light-scattering experiments of Law et
al.\cite{reva} in NESS of a one-component liquid, the scattering
wave number, $k$, was determined from equilibrium light-scattering
measurements and the known value of the thermal diffusivity, $D_T$.
In the present experiments we have determined $k$ by directly
measuring the scattering angle $\theta$. Since the intensities of
the nonequilibrium fluctuations depend on $k^{-4}$, the measurement
of $k$ has to be done with considerable accuracy. By working out
the ray tracing problem, taking into account refraction at the two
window surfaces, we can deduce the scattering angle, $\theta$, from
the location of the pinhole relative to that of the transmitted
beam For this calculation the thickness of the bottom window, the
distance from the window to the collecting plane, the refractive
index of the window and the refractive index, $n$, of the solution
are needed. It was assumed that the scattering volume is in the
middle of the cell, but since the cell is very thin the possible
corrections to this assumption are negligible. The distances were
measured with an accuracy of $\pm 0.05$ cm. The value 1.7660 of the
refractive index of the window at the wavelength $\lambda$ of the
incident light was obtained from the manufacturer. The refractive
index, $n$, of the polystyrene solutions as a function of the
polymer concentration at 25$^{\circ}$C was measured with a
thermostated Abb\'{e} refractometer as described by Zhang et
al.\cite{kaithesis}. For the polystyrene solutions considered in
the present paper the refractive index can be represented
by\cite{kaithesis}:
\begin{equation}
n(w) = 1.490~49 + 0.092~20 \cdot w + 0.025~56 \cdot w^2.
\label{eq:nn}
\end{equation}
The scattering wave number $k$ is related to the scattering angle
$\theta$ through the Bragg condition $k = {4 \pi n \over \lambda}
\sin({\theta/2})$, where $\lambda$ is the wavelength of the
incident light. Taken into account the experimental errors in the
different magnitudes relevant to the calculation, we were able to
determine $k$ with an accuracy of 1.5\%.

As discussed in section II, we want to investigate the
concentration fluctuations, which yield the dominant contribution
to the Rayleigh spectrum. Since typical decay times for this mode
at the small angles employed are always larger than 0.5 s, we are
interested in the experimental correlograms for times starting
around 10$^{-2}$ s. Since this value is well above the region where
photomultiplier (PM) afterpulsing effects are important, unlike in
previous work\cite{reva,approx}, cross-correlation was not
necessary for our measurements. This simplifies the experimental
arrangement. The light exiting through the pinhole is focused,
through corresponding optics, onto the field selecting pinhole in
front of a single PM. The signal from the PM and the corresponding
discriminator is analyzed with an ALV-5000 multiple tau correlator.

\section{Experimental Procedure and Experimental Results}

Rayleigh-scattering measurements were obtained for six different
solutions of polystyrene in toluene, with concentrations $w$ in
weight fraction ranging from 0.50\% to 4.00\%. For each solution,
the measurements were performed at three to five scattering angles,
ranging from 0.4$^{\circ}$ to 0.9$^{\circ}$, which corresponds to
scattering wave numbers $k$ ranging from 900 cm$^{-1}$ to 2000
cm$^{-1}$. These small angles cause the stray light from
window-surface scattering to be dominant, assuring that the
measurements are in the heterodyne regime. Hence, the light
scattered from the inner surfaces of the windows plays the role of
a local oscillator and provides the background with which our
signal, the light scattered from the polymer solution, is mixed.

Before starting the experiments, the light-scattering cell was
carefully cleaned by flushing the cell with pure toluene for at
least one hour. Next, a gentle stream of nitrogen gas was
continuously directed through the cell to dry the inner walls and
to remove dust particles. After this cleaning procedure the polymer
solution was introduced into the cell through a 0.5 $\mu$m
millipore Millex HPLC teflon filter. We have been careful to remove
bubbles from the light-scattering cell while we were filling it
with the polymer solution. This cleaning and filling procedure was
repeated each time the concentration of the solution inside the
cell was changed.

It should be noted that a portion of the same polymer solution was
introduced into an optical beam-bending cell for measuring the
diffusion coefficient, $D$, and the Soret coefficient, $S_T$, of
the solutions as described by Zhang et al.\cite{kaithesis}. These
independent measurements of the diffusion and Soret coefficients
will be used to compare the results of our light-scattering
measurements with the theoretical predictions. In the case of
polymer solutions it is especially important to have independent
measurements of these quantities for the same polymer/solvent
system because, as commented below, there is a sizable dispersion
in the literature data, mainly caused by the dependence of these
coefficients on parameters difficult to control, such as the
polydispersity of the sample.

\begin{figure}
\[
\psfig{file=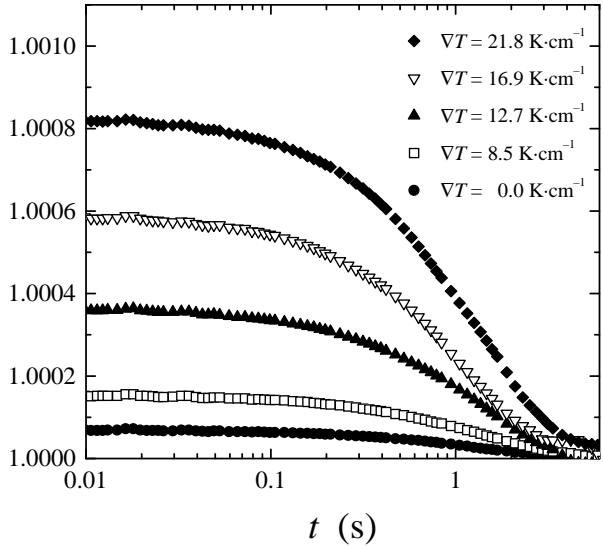,width=8.5cm}
\]
\caption{Normalized experimental light-scattering
correlation functions, obtained at $k=1030$ cm$^{-1}$ for a
solution of polystyrene ($M_W=96,400$, $w = 2.50\%$) in toluene
subjected to various temperature gradients, $\nabla T$, plotted
versus log($t$).}
\label{fnon2}
\end{figure}

Once the cell was filled with a polystyrene solution with known
polymer concentration, we started the experiments by setting the
temperature of the top and the bottom plates of the cell at
25$^{\circ}$C. A light-scattering angle was then selected with the
collection pinhole, and the distance in the collection plane
between the pinhole and the center of the transmitted forward-beam
spot was measured accurately. As already mentioned, this procedure,
with the corresponding calculations, yielded scattering wave
vectors $k$ with an accuracy of 1.5\%. Once a scattering angle was
selected, the optics was arranged to focus the light exiting
through the collecting pinhole into the field-stop pinhole at the
PM. When the temperature of the polymer solution had stabilized, we
used the ALV-5000 to collect at least ten equilibrium
light-scattering correlation functions, with the polymer solution
in thermal equilibrium at 25$^{\circ}$C. The photon count rate of
these measurements ranged from 0.3 to 2.5 MHz, depending on the
run. Each correlation lasted from 30 to 60 minutes, with signal to
background ratios from $1\times 10^{-4}$ to $1\times 10^{-3}$.
These small signal to background ratios confirm that our
measurements are in the heterodyne regime. After having completed
the measurements with the polymer solution in thermal equilibrium,
we applied various values of the temperature gradient to the
polymer solution by increasing and decreasing the temperatures of
the top and bottom plates symmetrically, so that all experimental
results correspond to the same average temperature of
25$^{\circ}$C. The maximum temperature difference employed was
4.1$^{\circ}$C, which corresponds to a maximum temperature gradient
of 34.6~K~cm$^{-1}$. The variation in the thermophysical properties
of the solutions is negligible in this small temperature interval,
and the property values corresponding to 25$^{\circ}$C have been
used for the calculations. After changing the temperatures, we
waited at least two hours, to be sure that the concentration
gradient was fully developed. Then, for each value of the
temperature gradient, about six correlograms were taken. In Fig.\
\ref{fnon2}, typical experimental light-scattering correlograms,
obtained with the ALV-5000 correlator at $k=1030$ cm$^{-1}$ are
shown for the solution with polymer mass fraction $w = 2.50\%$, as
a function of the temperature gradient $\nabla T$. The figure shows
the well sampled amplitude of the correlation at short times for
the correlograms.  A simple glance at Fig.\ \ref{fnon2} confirms
that the amplitude of the correlation function increases with
increasing values of the temperature gradient $\nabla T$.

The correlograms, in the range from 10$^{-2}$ s to 1 s, depending
on the run, were fitted to a single exponential, in accordance with
Eq. (\ref{eq:corr}). In all cases, both in equilibrium and in
nonequilibrium, very good fits were obtained. Actually, the
intensity of the scattered light is observed over a range of wave
numbers corresponding to the non-zero aperture of the collecting
pinhole. This effect can be accounted for by representing the
experimental correlation function in terms of a Gaussian
convolution, as explained in previous
publications\cite{reva,physica}. However, we found that the
resulting corrections to the parameter values obtained from fits to
a single exponential amounted to less than 1\% for the present
experiments and could be neglected. Hence, experimental values for
the decay rate, $Dk^2$, and for the prefactor $C_0[1+A_c(k,\nabla
T)]$ were obtained by directly fitting the time-correlation
function data of the scattered light to a single exponential, as
given by Eq. (\ref{eq:corr}).

The experimental values obtained for the decay rates, $Dk^2$, for
the polymer solutions at various values of the scattering wave
number $k$ and the temperature gradient $\nabla T$ are presented in
Table\ \ref{t1}. Each value displayed in Table\ \ref{t1} is the
average from several (at least five) experimental correlograms
obtained at the same conditions. The prefactor of the exponential
obtained from the fitting procedure was multiplied by the average
count rate during the run to get the average intensity (in 10$^6$
counts/s) in each run. The dimensionless enhancement, $A_c$, of the
concentration fluctuations at a given nonzero value of $\nabla T$
was obtained from the ratio of the nonequilibrium to the
equilibrium intensities, measured at the same $k$, by means of Eq.
(\ref{eq:corr}). The experimental values obtained for $A_c$ are
presented in Table\ \ref{t2}. Again, each value displayed in Table\
\ref{t2} is an average of several correlograms measured at the same
conditions. The uncertainties quoted in Tables\
\ref{t1}~and~\ref{t2} represent standard deviations of the values
obtained from the sets of experimental correlograms.

\section{Analysis of Experimental Results}
\subsection{Mass-diffusion coefficient $D$}

\begin{figure}
\[
\psfig{file=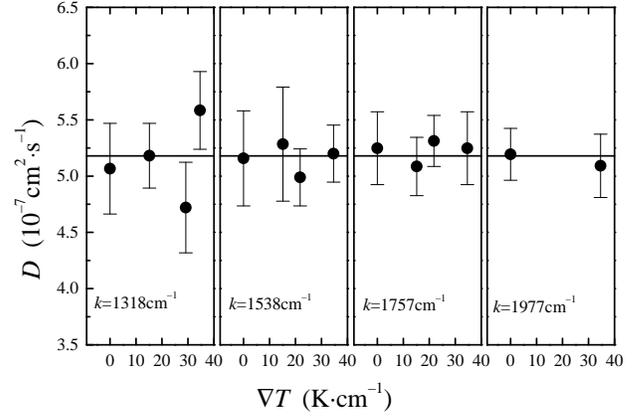,width=8.5cm}
\]
\caption{Collective mass-diffusion coefficient $D$ of polystyrene
($M_W=96,400$, $w=0.50\%$) in toluene deduced from the decay rates
of the concentration fluctuations at various values of $k$ and
$\nabla T$. The horizontal line represents the average value $D =
(5.19 \pm 0.04) \times 10^{-7}$ cm$^{2}$ s$^{-1}$.}
\label{f68}
\end{figure}

\begin{figure}
\[
\psfig{file=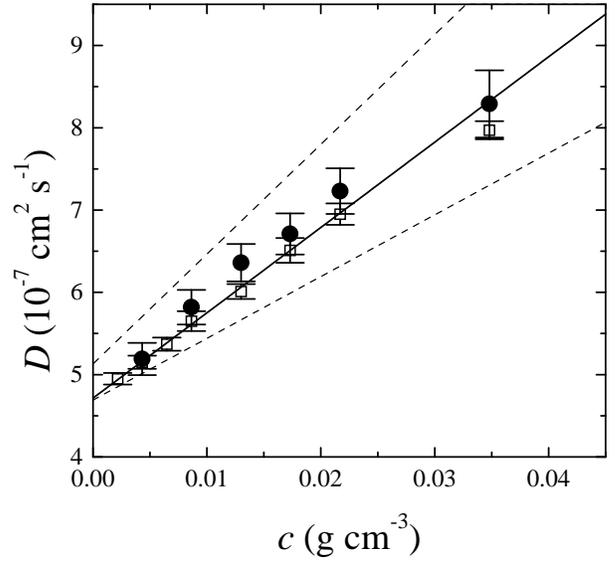,width=8.5cm}
\]
\caption{Collective mass-diffusion coefficient $D$ as a
function of concentration for polystyrene/toluene solutions
($M_W=96,400$). $\bullet$: measured in this work by equilibrium and
nonequilibrium Rayleigh light scattering. $\Box$: measured by Zhang
et al. al. with an optical beam-bending
technique\protect\cite{kaithesis}. The straight line represents Eq.
(\protect\ref{eq:dif}). The dashed lines represent the limits over
which the literature values are spread.}
\label{fD}
\end{figure}

Dividing the decay rates displayed in Table\ \ref{t1} by the square
of the known wave numbers, a mass-diffusion coefficient, $D$, is
obtained for each $w$, $k$ and $\nabla T$. As an example, we show
in Fig.\ \ref{f68} the values of $D$ thus obtained for the polymer
solution with $w=0.50\%$ obtained from the light-scattering
measurements at various values of the wave number $k$, as a
function of the applied temperature gradient $\nabla T$. The
horizontal line in Fig.\ \ref{f68} represents the weighted average
value of $D$. It is readily seen that the experimental $D$ is
independent of $\nabla T$, which implies that the applied
temperature gradient does not affect the translational diffusion
dynamics of the polymer molecules and that we indeed observed
concentration fluctuations for all $\nabla T$. Figure\ \ref{f68}
also demonstrates that $D$ is independent of the wave number $k$.
Thus the decay rates in Table\ \ref{t1} are indeed proportional to
$k^2$, which confirms the correctness of our measurements of $k$.
For each concentration we determined a simple value of $D$ as an
average of the experimental data obtained at various $k$ and
$\nabla T$. In the averaging process the individual accuracies were
taken into account. The resulting values of $D$ for the different
polymer solutions are presented in Table\ \ref{td} and displayed in
Fig.\ \ref{fD} as a function of the polymer concentration. For this
purpose we prefer to use the concentration $c$ in g cm$^{-3}$,
because this unit is more widely employed in the literature for
polymer solutions. To change the concentration units we used the
relationship:
\begin{equation}
\rho(w) = 0.86178 + 0.1794~w + 0.0296~w^2~({\rm g~cm^{-3}}),
\end{equation}
as reported by Scholte\cite{scholte}. The error bars in Fig.\
\ref{fD} have been calculated by adding 3\% of the value of $D$ to
the standard deviations given in Table\ \ref{td}, so as to account
for the uncertainty in the wave number $k$. In Fig.\ \ref{fD} we
have also plotted the values obtained by Zhang et al. for the
collective diffusion coefficient of the same polymer solutions with
an optical beam-bending technique\cite{kaithesis}. This figure
shows that the values obtained from the two methods for the
diffusion coefficient agree within the experimental accuracy. The
straight line displayed in Fig.\ \ref{fD} represents the linear
relationship:
\begin{equation}
D(c) = D_0(1+k_D~c),
\label{eq:dif}
\end{equation}
with the values
\begin{mathletters}
\label{eq:dvalues}
\begin{equation}
D_0 = (4.71\pm0.08)\times10^{-7}~{\rm cm}^2 {\rm s}^{-1},
\end{equation}
and
\begin{equation}
k_D = (22\pm2)~{\rm cm}^3 {\rm g}^{-1}. \label{equation}
\end{equation}
\end{mathletters}
for the diffusion coefficient at infinite dilution, $D_0$, and the
hydrodynamic interaction parameter, $k_D$, as determined by Zhang
et al.\cite{kaithesis} for the polymer solution with the same
molecular mass $M_W = 96,400$. It may be observed that the $D_0$
and $k_D$ values proposed by Zhang et al.\cite{kaithesis} yield a
satisfactory description of the dependence of our $D$ values on the
concentration.

An extensive survey of literature values for $D_0$ and $k_D$ of
polystyrene in toluene solutions was performed, and ten different
references were examined
\cite{kaithesis,w187,w188,w189,w190,w191,w192,w193,w168,soret}.
Most authors have studied the molecular weight dependence of these
parameters and propose relationships which usually have the form of
power laws. The results of our survey are presented in Table\
\ref{t3}. In some cases where the scaling equations are not
directly given by the authors, we have performed the corresponding
fits to obtain the power-law dependence of $D_0$ and $k_D$ on
$M_W$. All data in Table\ \ref{t3} correspond to polystyrene in
toluene at the temperature of 25$^{\circ}$C. While $k_D$ should be
nearly independent of temperature because toluene is a good
solvent\cite{w168}, $D_0$ depends on temperature through a
Stokes-Einstein relation. Neglecting the dependence on temperature
of the hydrodynamic radius of polystyrene in toluene\cite{w195}, we
represent the dependence of $D_0$ on the temperature, $T$, by:
\begin{equation}
D_0 \propto {T \over \eta_{0}(T)},
\label{eq:dependT}
\end{equation}
where $\eta_{0}(T)$ is the viscosity of the pure solvent (toluene)
as a function of temperature which can be found in the
literature\cite{w195}. Equation (\ref{eq:dependT}) was used for
making temperature corrections in cases where the values of $D_0$
reported in the literature had been measured at temperatures other
than 25$^{\circ}$C. Furthermore, we did not consider literature
values measured at temperatures more than $\pm$10 K from
25$^{\circ}$C. The sixth column of Table\ \ref{t3} contains the
values extrapolated to $M_W=96,400$, from the $D_0-M_W$ and
$k_D-M_W$ relationships shown in the second column of the same
table. From the information in Table\ \ref{t3} we conclude that
$D_0 = (4.91 \pm 0.22)\times10^{-7}$ cm$^2$ s$^{-1}$ and $k_D = 21
\pm 5$ cm$^3$ g$^{-1}$, to be compared with the values $D_0 = (4.71
\pm 0.08)\times10^{-7}$ cm$^2$ s$^{-1}$ and $k_D = 22 \pm 2$ cm$^3$
g$^{-1}$ quoted in Eq. (\ref{eq:dvalues}) and adopted by us. Note
that standard deviations of the extrapolated values for $D_0$ and
$k_D$ from the literature are $\pm 4.7\%$ and $\pm22\%$,
respectively. The corresponding spread of the literature values of
$D$ is indicated by the two dashed lines in Fig.\ \ref{fD}. The
upper line was calculated by taking for $D_0$ and $k_D$ the
literature averages plus their standard deviations; the lower line
was calculated by taking for $D_0$ and $k_D$ the literature
averages minus their standard deviations.

\subsection{Nonequilibrium enhancement $A_c$ of the
concentration fluctuations}

\begin{figure}
\[
\psfig{file=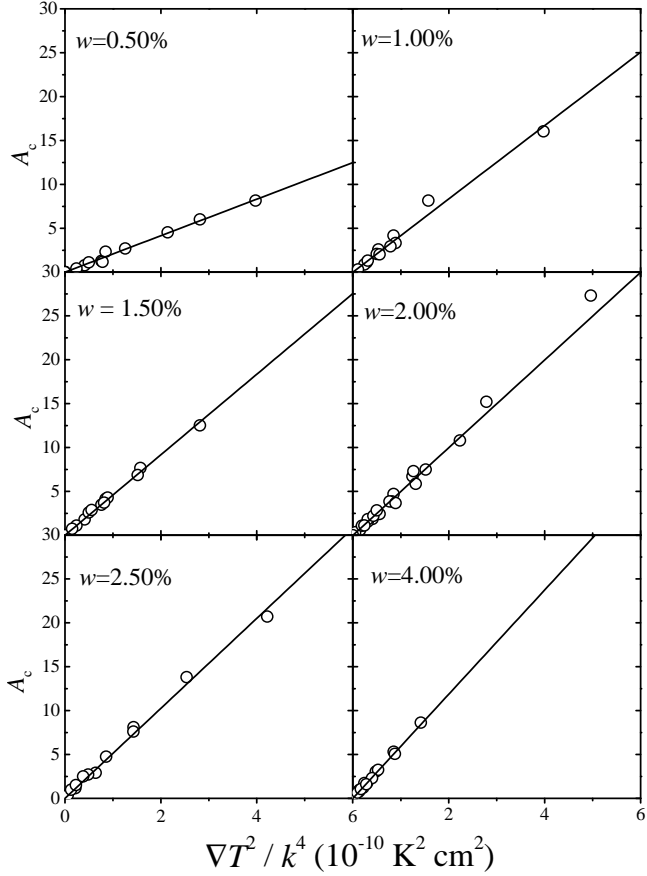,width=8.5cm}
\]
\caption{The amplitude $A_c$ of the nonequilibrium concentration fluctuations
in dilute and semidilute polystyrene/toluene solutions
($M_W=96,400$) at 25$^\circ$C and at various concentrations as a
function of $(\nabla T)^2/k^4$. The solid lines represent linear
fits to the experimental data for each concentration with slope
$A_c$ given in Table \protect\ref{ta}.}
\label{f2}
\end{figure}

Having confirmed the validity of our experimental results for the
decay rates, we now consider the nonequilibrium enhancements,
$A_c(\nabla T,k)$, of the concentration fluctuations reported in
Table\ \ref{t2}. As can be readily seen, the nonequilibrium
enhancements show a dramatic increase with increasing temperature
gradients. The experimental values obtained for the enhancement are
plotted in Fig.\ \ref{f2} as a function of $(\nabla T)^{2}/k^{4}$,
for the six polymer solutions investigated. The results of a
least-squares fit of the experimental points to a straight line
going through the origin are also displayed. The information
presented in Fig.\ \ref{f2} confirms that, in the range of
scattering wave vectors investigated, the nonequilibrium
enhancement of the concentration fluctuations is indeed
proportional to $(\nabla T)^{2} / k^{4}$ in accordance with Eq.
(\ref{eq:enhan1}).

The slope of the lines in Fig.\ \ref{f2} yields experimental values
for the strength of the enhancement, $A_{c}^{*}(w)$, listed in
Table\ \ref{ta}. To compare the experimental results with the
theoretical prediction, Eq. (\ref{eq:enhan2}), we need several
thermophysical properties of the solutions, namely: the
concentration derivative of the difference in the chemical
potentials per unit mass, $({\partial \mu / \partial w})_{p,T}$,
the zero-shear viscosity, $\eta$, the mass-diffusion coefficient,
$D$, and the Soret coefficient, $S_T$.

{\em a)} The derivative of the difference in chemical potentials
was calculated from its relationship with the osmotic
pressure\cite{mu}, $\Pi$, which, for the small concentrations used
can be simplified to:
\begin{equation}
\left( {\partial \mu \over \partial w} \right)_{P,T} =
{1 \over \rho w} ~
\left( {\partial \Pi \over \partial w} \right)_{P,T},
\label{eq:mu}
\end{equation}
Values for the concentration derivative of the osmotic pressure of
the solutions, were obtained from the extensive work of Noda et
al.\cite{noda1,noda2}. Specifically, the universal function:
\begin{equation}
\left( {\partial \Pi \over \partial c} \right)_{P,T}=
{RT \over M_W} \left[ 1 + 2 \left({3 \sqrt{\pi} \over 4}~{c \over
c^*} \right) + {3 \over 4}~\left({3 \sqrt{\pi} \over 4}~{c \over
c^*} \right)^2 \right]
\end{equation}
represents the behavior of this property from the very dilute to
the concentrated regime. The overlap concentration, $c^* = 0.0272$
g cm$^{-3}$, was calculated, as in section III, from the
correlations proposed by the same authors\cite{noda1,noda2}.

{\em b)} To calculate the zero-shear viscosity, we employed the
Huggins relationship:
\begin{equation}
\eta = \eta_0 (1 + [\eta] c + k_{\rm H} [\eta]^2 c^2),
\label{eq:nu}
\end{equation}
where $\eta_0$ is the viscosity of the pure solvent, for which the
value 552.7 mPa$\cdot$s was taken, $[\eta]$ is the intrinsic
viscosity of polystyrene in toluene and $k_{\rm H}$ the Huggins
coefficient for the same system. The intrinsic viscosity was
obtained from the correlation\cite{noda1,huggins1}: $[\eta] =
9.06~10^{-3} M_W^{0.74}~({\rm cm^3~g^{-1}})$. For the Huggins
coefficient, the usual value in a good
solvent\cite{huggins1,huggins2}, $k_{\rm H} = 0.35$, was adopted;
no molecular weight dependence has been reported in the literature
for $k_{\rm H}$.

{\em c)} Values for the collective mass-diffusion coefficient, $D$,
of our solutions were presented in the previous section, when
analyzing the decay rates of the correlograms. They are displayed
in Table\ \ref{td}. For a continuous representation of $D$ as a
function of concentration we use Eq. (\ref{eq:dif}), with the
parameters given by Eq. (\ref{eq:dvalues}).

{\em d)} The Soret coefficient, $S_T$, was measured by Zhang et
al.\cite{kaithesis} for the same solutions used in our
light-scattering experiments. It is worth noting that the Soret
coefficients measured by Zhang et al. agree, within experimental
error, with other recent $S_T$ values for polystyrene in toluene
reported in the literature\cite{soret}. Since the strength of the
enhancement depends on the square of the Soret coefficient, we need
a good continuous representation of these data to make a
theoretical prediction of $A_c^*$ as a function of the
concentration. We assume that $S_T$ scales as the inverse of $D$,
as rationalized by Brochard and de Gennes\cite{degennes}, and use a
relationship proposed by Nystrom and Roots\cite{nystrom}and used
successfully in Zhang, et al.\cite{kaithesis} for the diffusion
coefficient and Soret coefficient.  The equation\cite{note1} for
$S_T$ is
\begin{equation}
S_T = S_{T0} {(1+X_S)^{A} \over 1 + A_S X_S (1+X_S)^{B}},
\label{eq:nystrom}
\end{equation}
where   $A = {(1-\nu)\over (3\nu -1)}$ and $B = {(2-3\nu)\over
(3\nu -1)}$ are exponents evaluated with $\nu =0.588$, $X_S=r_S(k_S
c)$ is a scaling variable, and $A_S = A + r_S ^{-1}$. The virial
constants $S_{T0}$ and $k_S$ are defined  by making  a series
expansion of Eq. (\ref{eq:nystrom}) around $c=0$ to give $S_T =
S_{T0} (1 - k_S c) + \cdot\cdot\cdot$, where $k_S$ is the first
virial coefficient of $S_T$.

\begin{figure}
\[
\psfig{file=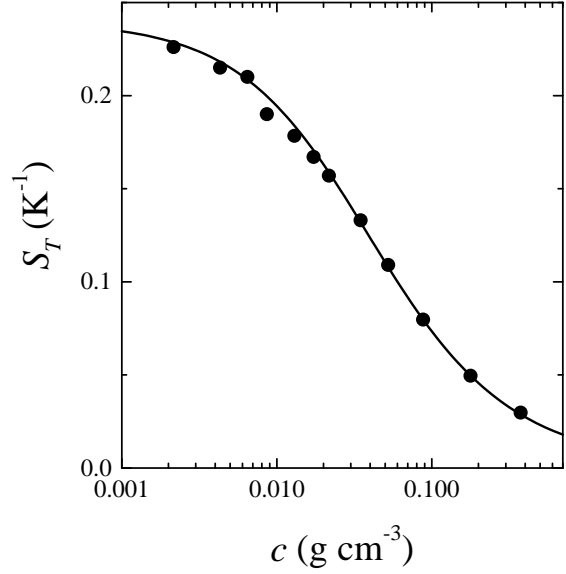,width=8.5cm}
\]
\caption{Experimental values of the Soret coefficient of
polystyrene ($M_W = 96,400$) in toluene at 25$^{\circ}$ as a
function of polymer concentration $c$. The solid curve represents
Eq. (\protect\ref{eq:nystrom}) with $k_S= 24$ cm$^3$ g$^{-1}$,
$S_{T0}= 0.24$ K$^{-1}$, and $r_S$ determined by fitting to be
$r_S=1.16 \pm 0.07$
.}
\label{fsoret}
\end{figure}

We used the values of $k_S= 24$ cm$^3$ g$^{-1}$ and $S_{T0}= 0.24$
K$^{-1}$ found by Zhang et al.\cite{kaithesis} for this molecular
weight and then did a weighted, least-squares fit to the
concentration dependent $S_T$ data of Zhang et al. to find a value
of $r_S$. We find $r_S=1.16 \pm 0.07$.  Equation (\ref{eq:nystrom})
gives an excellent representation of the experimental Soret
coefficient data, as shown in Fig.\ \ref{fsoret}.

\begin{figure}
\[
\psfig{file=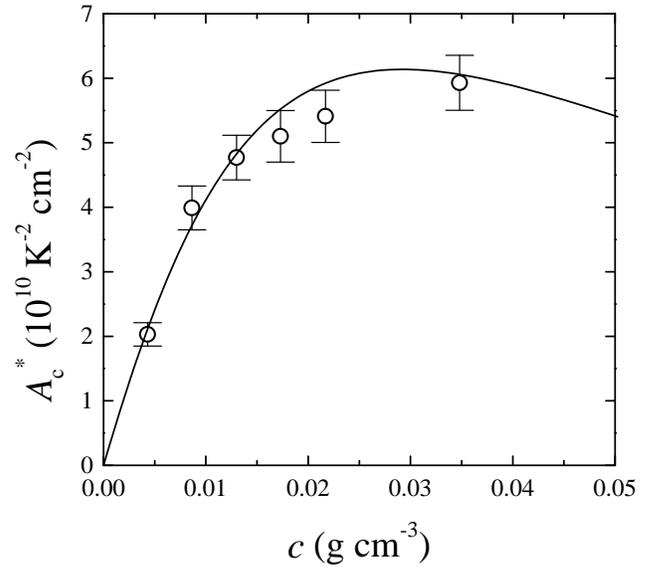,width=8.5cm}
\]
\caption{Comparison of the measured values of the strength, $A_c^*$,
of the nonequilibrium enhancement of the concentration fluctuations
(open circles) with the theoretical values (solid curve) calculated
from Eq. (\protect\ref{eq:enhan2}).}
\label{f3}
\end{figure}

In Fig.\ \ref{f3} we present a comparison between the experimental
values for the nonequilibrium-enhancement strength, $A_c^*$, and
the values calculated form Eq. (\ref{eq:enhan2}) with the
information for the various thermophysical properties as specified
above. The error bars associated with the experimental data
displayed in Fig.\ \ref{f3} have been calculated by adding 6\% to
the statistical errors quoted in Table\ \ref{ta} so as to account
for a 1.5\% uncertainty in the values of the wave number $k$. The
theoretical values have an estimated uncertainty of at least 5\%.
Taking into account these uncertainties, we conclude that the
observed strength of the nonequilibrium concentration fluctuations
is in agreement with the values predicted on the basis of
fluctuating hydrodynamics in the concentration range investigated.

\section{Conclusions}

The existence of long-range concentration fluctuations in polymer
solutions subjected to stationary temperature gradients has been
verified experimentally. As in the case of liquid
mixtures\cite{physica,approx,cell}, the nonequilibrium enhancement
of the concentration fluctuations has been found to be proportional
to $(\nabla T)^2/k^4$.

Unlike the case of liquid mixtures\cite{physica,approx,cell}, good
agreement between the experimental and theoretical values for the
strength of the enhancement of the concentration fluctuations,
$A_c^*$, has been found here. This indicates the validity of
fluctuating hydrodynamics to describe nonequilibrium concentration
fluctuations in dilute and semidilute polymer solutions. Further
theoretical and experimental work will be needed to understand
nonequilibrium concentration fluctuations in polymer solutions at
higher concentrations.

Our present results complement the considerable theoretical and
experimental progress recently made in understanding the dynamics
of concentration fluctuations of polymer solutions under shear
flow. It has been demonstrated that concentration fluctuations in
semidilute polymer solutions subjected to shear flow are also
enhanced dramatically\cite{shear1,shear2,shear3}. This enhancement
of the intensity of the concentration fluctuations in shear-induced
NESS, is similar to the enhancement reported here, when the NESS is
achieved by the application of a stationary temperature gradient.
On the other hand, the long-range nature of the concentration
fluctuations when a concentration gradient is present, has also
recently been observed in a liquid mixture by shadowgraph
techniques\cite{nature}, yielding additional evidence of the
interesting nature of this topic.

\section*{Acknowledgements}

We are indebted to J.F.~Douglas for valuable discussions and to
S.C.~Greer for helpful advice concerning the characterization of
the polymer sample. J.V.S. acknowledges the hospitality of the
Institute for Theoretical Physics of the Utrech University, where
part of the manuscript was prepared. The research at the University
of Maryland is supported by the U.S. National Science Foundation
under Grant CHE-9805260. J.M.O.Z. was funded by the Spanish
Department of Education during his postdoctoral stage at Maryland,
when part of the work was done.

\newpage
\mediumtext
\begin{tighten}
\begin{table}
\squeezetable
\caption{Experimental decay rates $Dk^2$ (in s$^{-1}$),
measured at different concentrations $w$, temperature gradients
$\nabla T$ and scattering wave numbers $k$}
\begin{tabular}{cccccccccc}

$w$&$k$&$\nabla T$=0.0&$\nabla T$=8.5&$\nabla T$=12.7&$\nabla
T$=15.8&$\nabla T$=16.9 &$\nabla T$=21.8&$\nabla T$=29.1&$\nabla
T$=34.6\\ &(cm$^{-1}$)&(K cm$^{-1}$)&(K cm$^{-1}$)&(K cm$^{-1}$)&(K
cm$^{-1}$) &(K cm$^{-1}$)&(K cm$^{-1}$)&(K cm$^{-1}$)\\
\tableline

0.50\%&1318&0.88$\pm$0.07&&&0.90$\pm$0.05&&&0.82$\pm$0.07&0.97$\pm$0.06\\
&1538&1.22$\pm$0.10&&&1.25$\pm$0.12&&&1.18$\pm$0.06&1.23$\pm$0.06\\
&1757&1.62$\pm$0.10&&&1.57$\pm$0.08&&&1.64$\pm$0.07&1.62$\pm$0.04\\
&1977&2.03$\pm$0.09&&&&&&&1.99$\pm$0.11\\

1.00\%&1318&1.00$\pm$0.06&&0.98$\pm$0.07&&&&0.98$\pm$0.07&1.00$\pm$0.06\\
&1538&1.4\ \ $\pm$0.3\ \ &&1.34$\pm$0.06&&&1.34$\pm$0.05\\
&1757&1.84$\pm$0.08&&&1.85$\pm$0.09&&1.87$\pm$0.08&1.80$\pm$0.12\\
&1977&2.28$\pm$0.10&&2.30$\pm$0.12&&&2.28$\pm$0.09&2.21$\pm$0.14&
2.34$\pm$0.12\\

1.50\%&1318&1.09$\pm$0.10&&&1.13$\pm$0.07&&1.12$\pm$0.04&1.14$\pm$0.06\\
&1538&1.46$\pm$0.09&&&1.45$\pm$0.08&&1.42$\pm$0.15&1.49$\pm$0.08\\
&1757&1.99$\pm$0.12&&&1.99$\pm$0.11&&1.96$\pm$0.09&2.03$\pm$0.10\\
&1977&2.49$\pm$0.14&&&2.54$\pm$0.14&&&2.47$\pm$0.09&2.44$\pm$0.12\\

2.00\%&872&0.52$\pm$0.06&0.49$\pm$0.03&0.46$\pm$0.02&&0.50$\pm$0.02\\
&1392&1.29$\pm$0.07&1.32$\pm$0.03&1.28$\pm$0.06&&1.18$\pm$0.08&
1.22$\pm$0.14\\
&1538&1.60$\pm$0.24&&&1.64$\pm$0.10&&1.62$\pm$0.08&1.62$\pm$0.06&
1.66$\pm$0.11\\
&1757&2.09$\pm$0.11&&&2.11$\pm$0.10&&2.14$\pm$0.07&2.06$\pm$0.10&
2.24$\pm$0.14\\
&1977&2.58$\pm$0.10&&&2.57$\pm$0.08&&2.63$\pm$0.08&2.60$\pm$0.14&
2.72$\pm$0.11\\

2.50\%&1030&0.75$\pm$0.04&0.72$\pm$0.04&0.75$\pm$0.02&&0.78$\pm$0.06&
0.73$\pm$0.06\\
&1350&1.43$\pm$0.15&1.32$\pm$0.05&1.35$\pm$0.11&&1.30$\pm$0.04&
1.33$\pm$0.07\\
&1880&2.66$\pm$0.07&2.50$\pm$0.11&2.67$\pm$0.12&&2.68$\pm$0.14&
2.53$\pm$0.12\\

4.00\%&1353&1.61$\pm$0.16&1.59$\pm$0.09&1.50$\pm$0.07&&1.49$\pm$0.07&
1.57$\pm$0.09\\
&1525&1.92$\pm$0.14&1.91$\pm$0.13&1.91$\pm$0.09&&1.92$\pm$0.15&
1.99$\pm$0.08\\
&1854&2.86$\pm$0.10&2.96$\pm$0.10&3.11$\pm$0.18&&3.07$\pm$0.08&
2.79$\pm$0.18\\
&2025&3.42$\pm$0.15&3.31$\pm$0.20&3.41$\pm$0.13&&3.23$\pm$0.05&
3.30$\pm$0.10\\

\end{tabular}
\label{t1}
\end{table}
\vfill
\begin{table}
\squeezetable
\caption{Experimental nonequilibrium enhancements, $A_c$,
measured at different concentrations $w$, temperature gradients
$\nabla T$ and scattering wave numbers $k$}
\begin{tabular}{ccccccccc}

$w$&$k$&$\nabla T$=8.5&$\nabla T$=12.7&$\nabla T$=15.8&$\nabla
T$=16.9&$\nabla T$=21.8& $\nabla T$=29.1&$\nabla T$=34.6\\
&(cm$^{-1}$)&(K cm$^{-1}$)&(K cm$^{-1}$)&(K cm$^{-1}$)&(K
cm$^{-1}$) &(K cm$^{-1}$)&(K cm$^{-1}$)&(K cm$^{-1}$)\\
\tableline

0.50\%&1318&&&1.28$\pm$0.17&&&6.0$\pm$0.5&8.12$\pm$0.3\\
&1538&&&0.78$\pm$0.11&&2.32$\pm$0.18&&4.53$\pm$0.15\\
&1757&&&0.41$\pm$0.04&&1.11$\pm$0.06&&2.69$\pm$0.07\\
&1977&&&&&&&1.17$\pm$0.07\\

1.00\%&1318&&2.58$\pm$0.10&&&8.1$\pm$0.7&&16.0$\pm$1.0\\
&1538&&1.10$\pm$0.19&&&4.2$\pm$0.3\\
&1757&&&0.87$\pm$0.08&&2.07$\pm$0.08&3.31$\pm$0.13\\
&1977&&0.30$\pm$0.07&&&1.30$\pm$0.05&2.02$\pm$0.14&2.94$\pm$0.08\\

1.50\%&1318&&&3.47$\pm$0.17&&7.7$\pm$0.4&12.5$\pm$0.8\\
&1538&&&1.80$\pm$0.11&&4.4$\pm$0.3&6.8$\pm$0.4\\
&1757&&&1.10$\pm$0.09&&2.59$\pm$0.10&4.29$\pm$0.12\\
&1977&&&0.73$\pm$0.07&&&2.87$\pm$0.12&3.71$\pm$0.19\\

2.00\%&872&6.7$\pm$0.4&15.2$\pm$0.3&&27.3$\pm$1.2\\
&1392&1.08$\pm$0.08&2.20$\pm$0.12&&3.84$\pm$0.22&7.3$\pm$0.3\\
&1538&&&1.86$\pm$0.13&&4.70$\pm$0.24&7.5$\pm$0.3&10.8$\pm$0.7\\
&1757&&&1.11$\pm$0.07&&2.82$\pm$0.24&3.66$\pm$0.25&5.85$\pm$0.35\\
&1977&&&0.59$\pm$0.04&&1.55$\pm$0.08&2.41$\pm$0.11&3.85$\pm$0.10\\

2.50\%&1030&2.9$\pm$0.3&8.1$\pm$0.5&&13.8$\pm$0.8&20.7$\pm$2.1\\
&1350&1.16$\pm$0.06&2.71$\pm$0.15&&4.74$\pm$0.11&7.6$\pm$0.3\\
&1880&0.37$\pm$0.08&0.97$\pm$0.15&&1.5$\pm$0.4&2.5$\pm$0.3\\

4.00\%&1353&1.20$\pm$0.17&3.04$\pm$0.17&&5.3$\pm$0.3&8.61$\pm$0.22\\
&1525&0.78$\pm$0.04&1.74$\pm$0.08&&3.24$\pm$0.19&5.06$\pm$0.21\\
&1854&0.48$\pm$0.09&0.94$\pm$0.08&&1.73$\pm$0.14&2.30$\pm$0.11\\
&2025&0.24$\pm$0.03&0.64$\pm$0.12&&1.05$\pm$0.04&1.59$\pm$0.04\\

\end{tabular}
\label{t2}
\end{table}
\end{tighten}

\newpage
\narrowtext
\begin{table}
\caption{Collective mass-diffusion coefficient measured
by equilibrium and nonequilibrium light-scattering of polystyrene
($M_W = 96400$) in toluene at 25$^{\circ}$C.}
\begin{tabular}{ccc}

$w$ (\%)&$c$ (g cm$^{-3}$)&$D$ (10$^{-7}$ cm$^{-2}$ s$^{-1}$)\\
\tableline 0.50&0.00431&5.19$\pm$0.04\\
1.00&0.00864&5.83$\pm$0.03\\ 1.50&0.0130&6.36$\pm$0.04\\
2.00&0.0173&6.71$\pm$0.05\\ 2.50&0.0217&7.23$\pm$0.06\\
4.00&0.0348&8.29$\pm$0.16\\
\end{tabular}
\label{td}
\end{table}

\newpage
\mediumtext

\begin{table}
\caption{Mass-diffusion coefficient at infinite dilution
($D_0$, in cm$^2$ s$^{-1}$) and hydrodynamic interaction parameter
($k_D$, in cm$^3$ g$^{-1}$) of polystyrene in toluene at
25$^{\circ}$C, obtained from various literature sources.}
\psfig{file=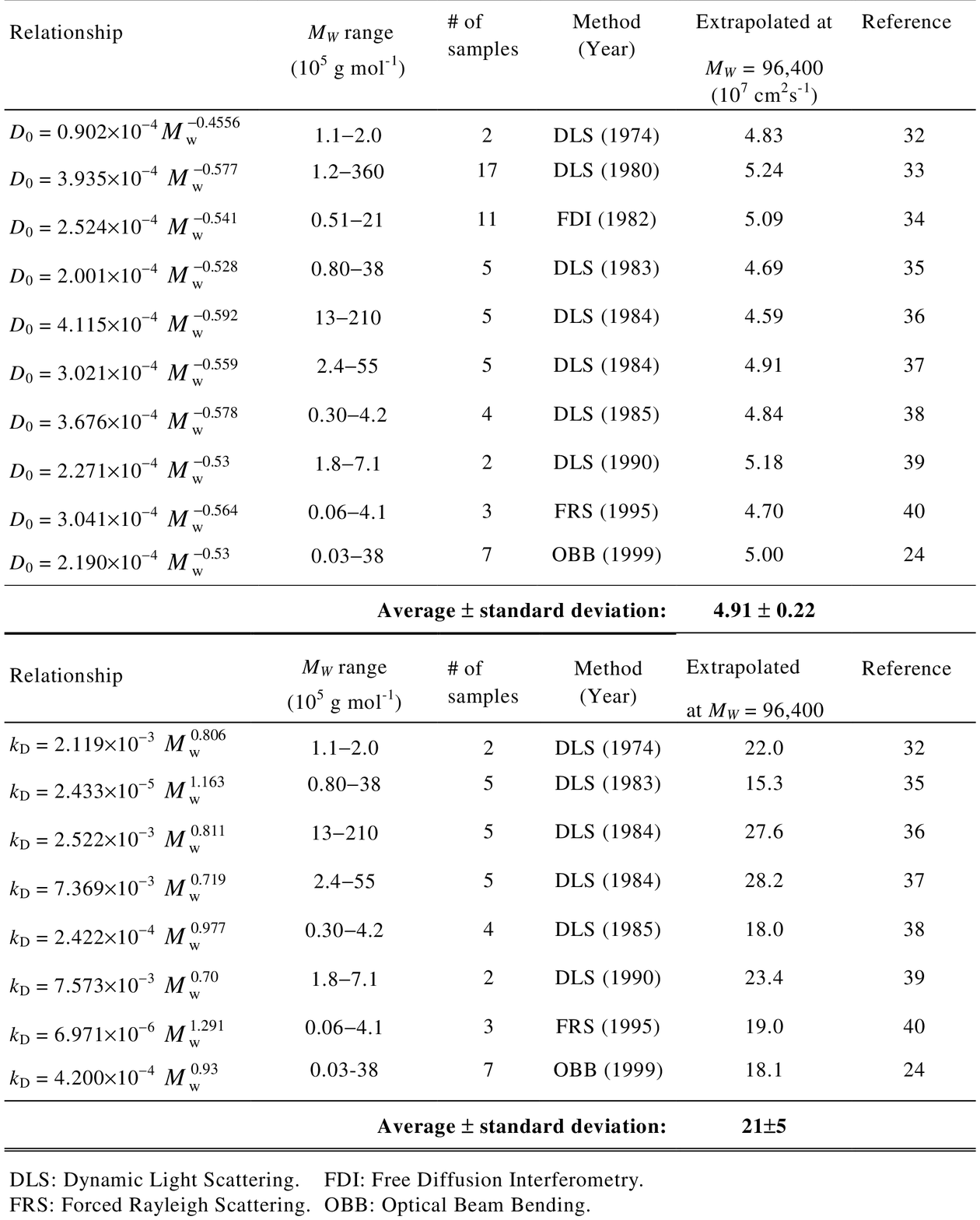,width=14cm}
\label{t3}
\end{table}

\newpage
\narrowtext

\begin{table}
\caption{Observed strength of the nonequilibrium
enhancement of the concentration fluctuations in solutions of
polystyrene in toluene at 25$^{\circ}$C.}

\begin{tabular}{ccc}

$w$ (\%)&$c$ (g cm$^{-3}$)&$A_c^*$ (10$^{10}$ K$^{-2}$ cm$^{-2}$)\\
\tableline

0.50&0.00431&2.03$\pm$0.07\\ 1.00&0.00864&4.00$\pm$0.10\\
1.50&0.0130&4.77$\pm$0.06\\ 2.00&0.0173&5.10$\pm$0.12\\
2.50&0.0217&5.41$\pm$0.08\\ 4.00&0.0348&5.93$\pm$0.07\\
\end{tabular}
\label{ta}
\end{table}

\newpage

\
\vfill    
\
\vfill

\end{multicols}

\begin{thebibliography}{99}
\bibitem{note1} The expression for the exponent $A$ given in
reference [\ref{lbkai2}] is given with the wrong sign.
\end{thebibliography}
\end{document}